\documentclass[11pt,twoside]{article}
\usepackage{asp2006}
\usepackage{psfig}
\usepackage{epsf}
\usepackage{graphics}
\usepackage{lscape}
\def\edcomment#1{\iffalse\marginpar{\raggedright\sl#1\/}\else\relax\fi}
\reversemarginpar

\begin{document}
\title{Spitzer spectra of Seyfert galaxies}
\author{Luigi Spinoglio \& Silvia Tommasin}
\affil{Istituto di Fisica dello Spazio Interplanetario,
  INAF, Via Fosso del Cavaliere 100, I-00133, Roma,
  Italy}
\author{Matthew A. Malkan \& Kevin Hainline}
\affil{Astronomy Division, University of California, Los Angeles, CA 90095-1547, USA}

\begin{abstract}
The Spitzer IRS high resolution spectra of about 90 Seyfert galaxies from the 12$\mu$m
Galaxy Sample are presented and discussed. These represent about 70\% of the total
complete sample of local Seyfert galaxies. The presence of starburst components in
these galaxies can be quantified by powerful mid-IR diagnostics tools (i.e. 11.25$\mu$m
 PAH feature equivalent width and the H$_2$ emission line intensity) as well as
 the AGN dominance can be measured by specific fine structure line ratios (e.g.
 [NeV]/[NeII], [NeV]/[SiII], etc.). The observed line ratios are compared to the results
 of semianalytical models, which can be used to compute the AGN and starburst
 contributions to the total luminosity of the galaxies. The results are also discussed 
 in the light of unification and evolution models.
\end{abstract}

\vspace{-0.5cm}

\section{Mid-IR spectroscopy with \textit{Spitzer} }

We briefly describe here some of the results of the mid-IR spectroscopic survey of Seyfert galaxies
belonging to the 12$\mu$m Galaxy sample (hereafter 12MGS)(Rush, Malkan \& Spinoglio 1993). 

The first results of the \textit{Spitzer} spectroscopic survey of the Seyfert galaxies of the 12$\mu$m 
sample \citep{t08} show a clear inverse trend between the indicator of 
\textit{AGN dominance}, the [NeV]14.3$\mu$m/[NeII]12.8$\mu$m line ratio, 
and the equivalent width of the 11.25$\mu$m PAH feature, which can be 
considered as an indicator of the \textit{star formation dominance}, as shown in Figure 1a, where the sample objects
increaed from 30 \citep{t08} to 87 \citep{t09}.
This result confirms an early finding of the ISO-SWS spectrometer \citep{ge98} with a much better statistics.
Here the Seyfert galaxies have been reclassified, following the results of spectropolarimetry \citet{tra1,tra3}, 
in type 1's (including the classical Seyfert 1's and the hidden Broad Line Region Seyfert 2's, as
discovered through spectropolarimetry) and ''pure" type 2's (for which a BLR was not detected). 
Most of the type 1 objects, 
including both Seyfert 1s and hidden Broad Line Region Seyfert 2s, are
located at high values of the [NeV]14.3$\mu$m/[NeII]12.8$\mu$m line ratio
and very low or absent PAH emission.

\begin{figure}[!h]
  \plottwo{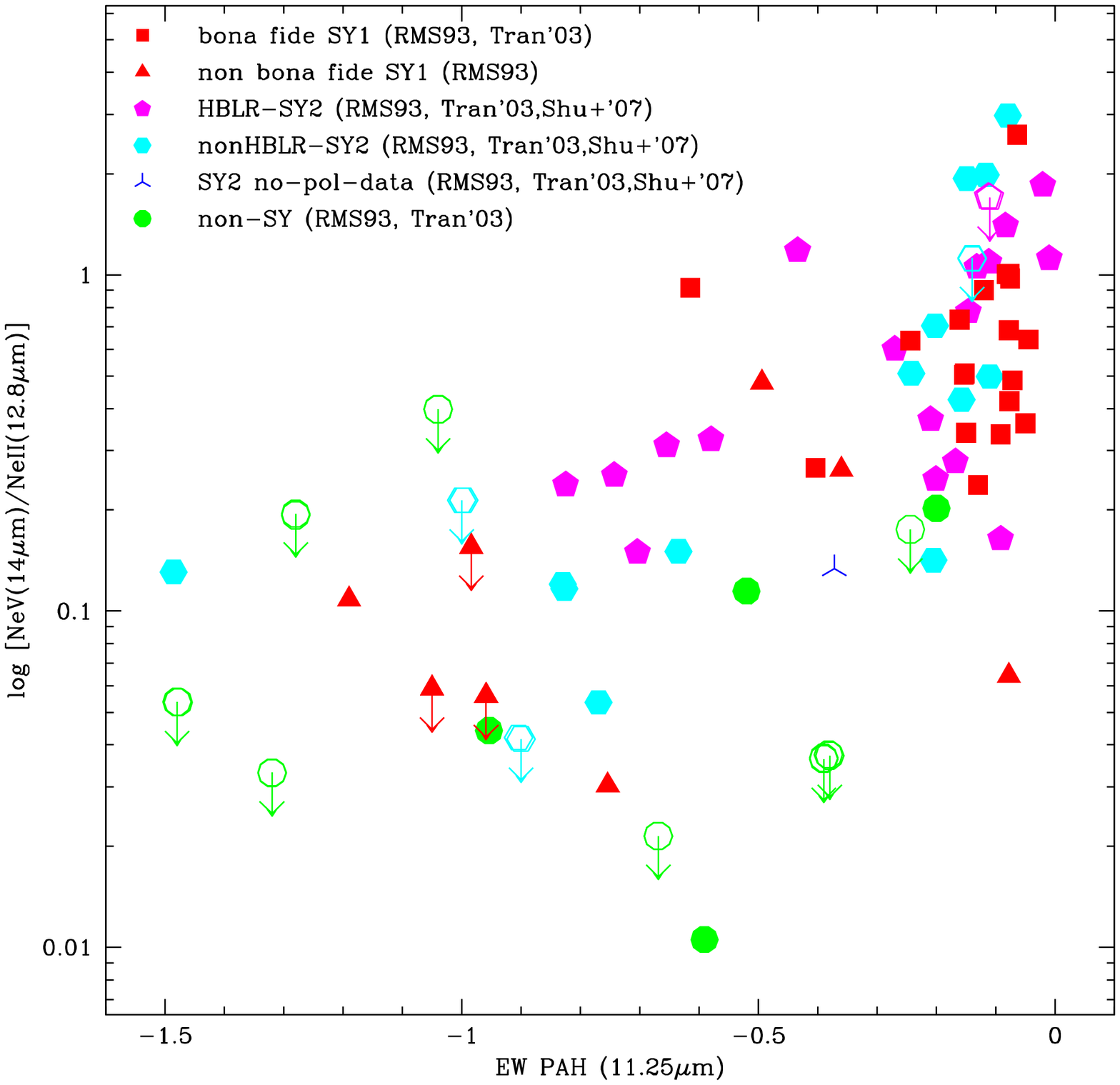}{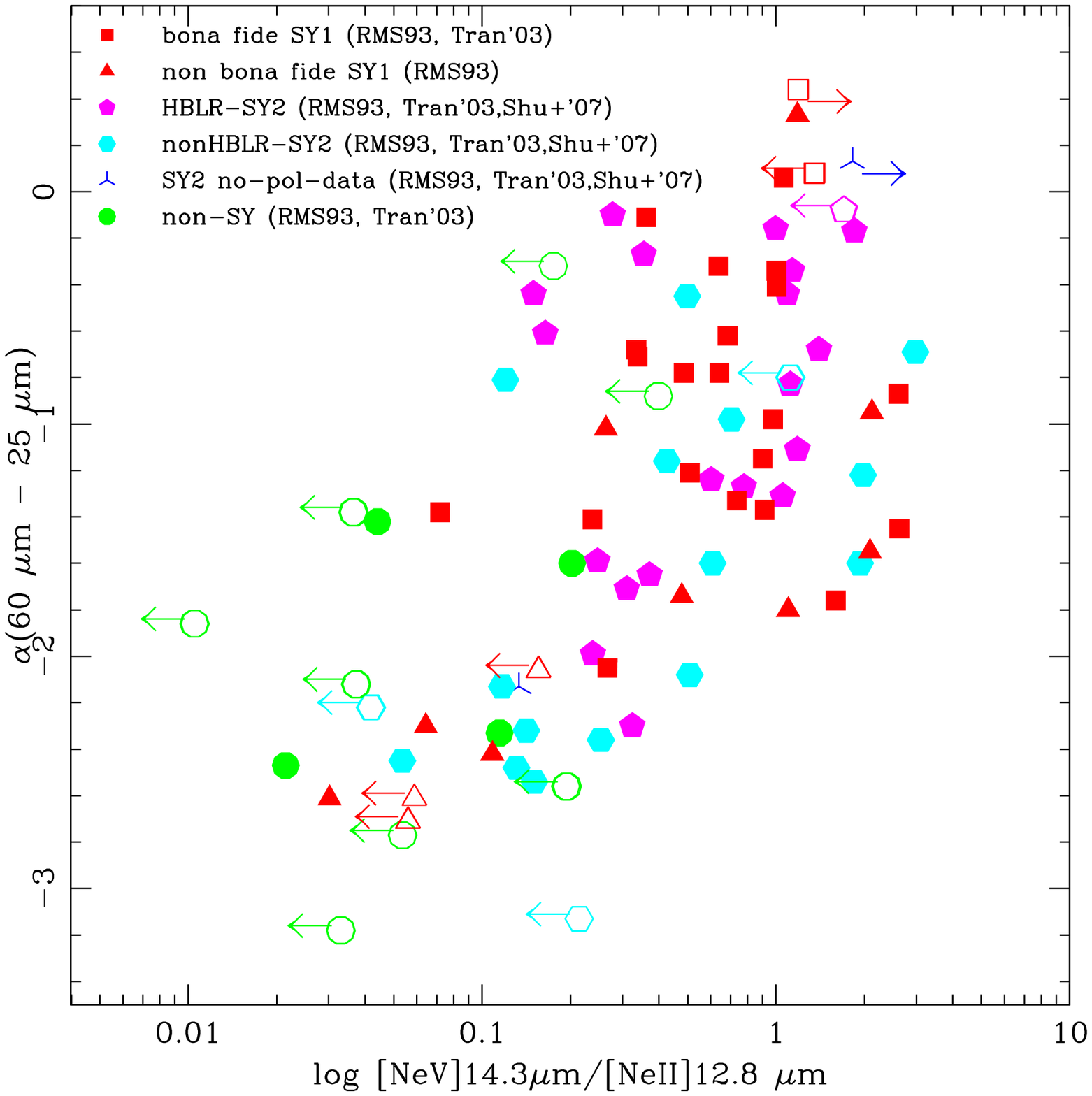}
  \caption{ {\bf a:(left)} [NeV]14.3$\mu$m/[NeII]12.8$\mu$m line ratio 
versus the equivalent width of the 11.25$\mu$m PAH. {\bf b:(right)} The mid-to-far-IR spectral 
index $\alpha_{(60\mu m-25\mu m)}$ versus the [NeV]14.3$\mu$m/[NeII]12.8$\mu$m line ratio \citep{t08,t09}.}
  \label{fig:neon_pah}
\end{figure}

Another diagnostic diagram using both spectroscopic and photometric results is shown in Figure 1b: 
the spectral index between 25 and 60$\mu$m $\alpha_{(60\mu m-25\mu m)}$ versus the 
[NeV]14.3$\mu$m/[NeII]12.8$\mu$m line ratio. A clear trend shows that when the  
\textit{AGN dominance} increases, the spectral index flattens. Most of type 1 objects
appear to be concentrated in the upper right part of the diagram, at high values of 
\textit{AGN dominance} and flat mid-to-far-IR slopes.

\begin{figure}[!h]
  \plottwo{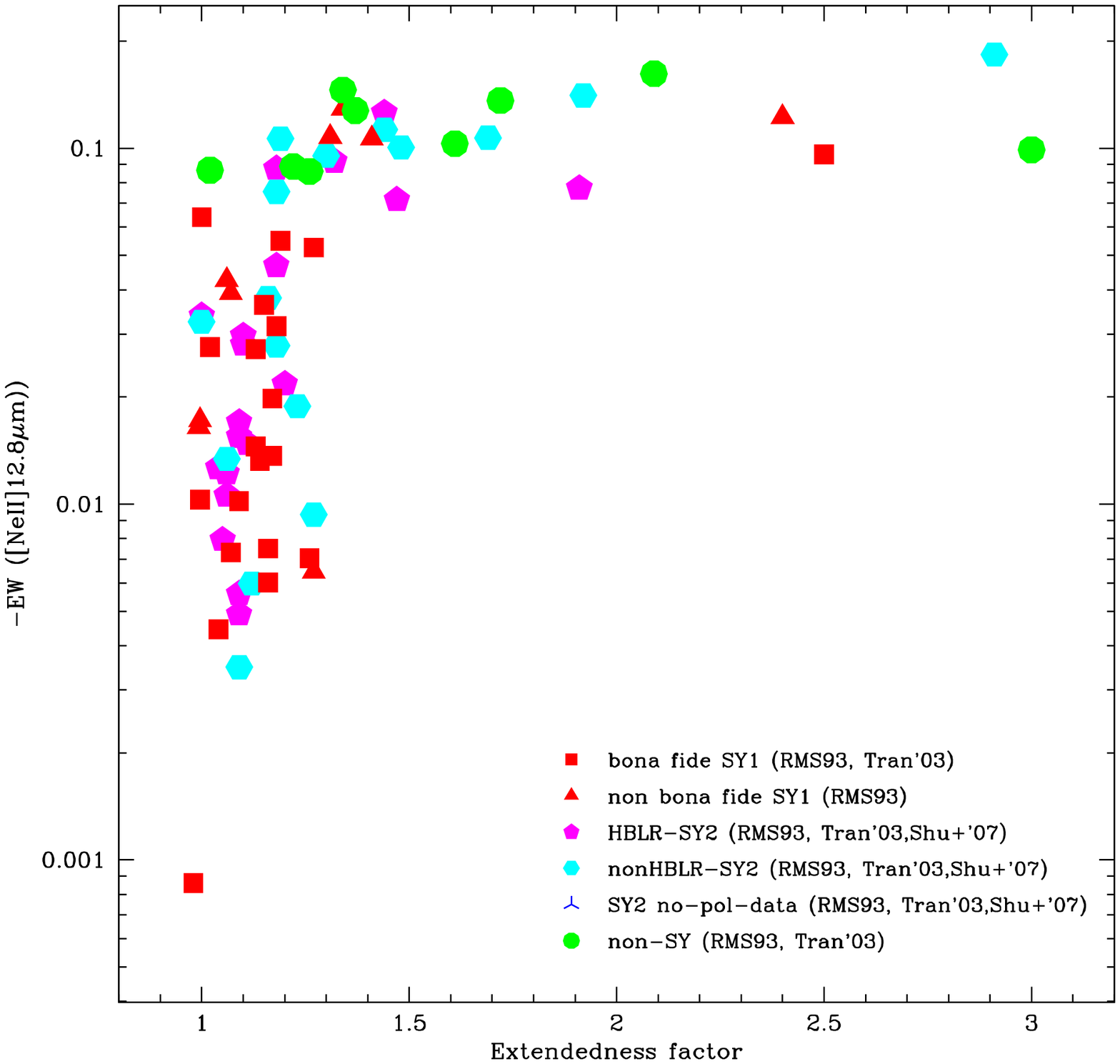}{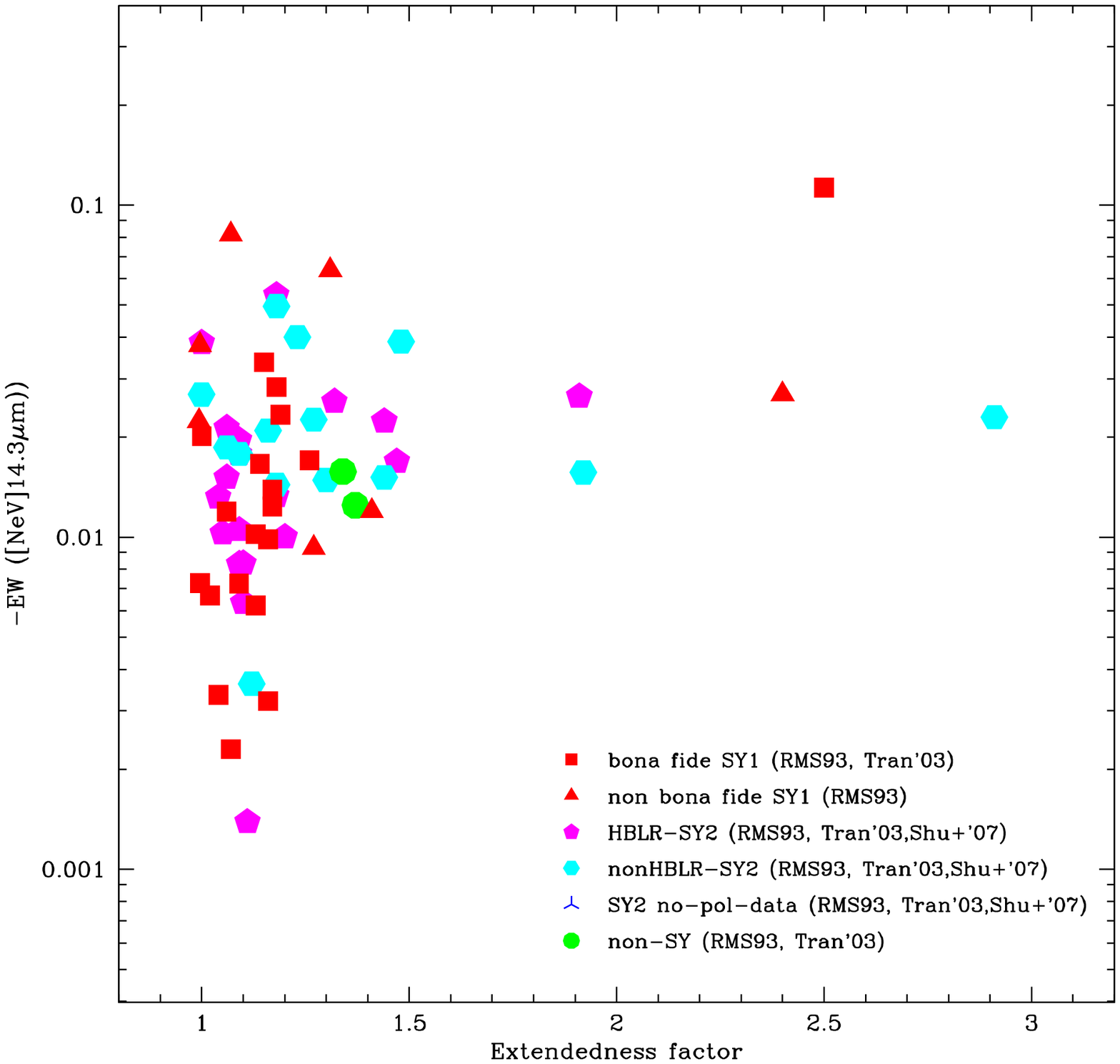}
  \caption{ {\bf a:(left)}  [NeII]12.8$\mu$m line equivalent width as a function of the source extendedness. 
  {\bf b:(right)} [NeV]14.3$\mu$m line EW versus source extendedness \citep{t08,t09}.}
  \label{fig:ext_ne2}
\end{figure}

The two channels of the Spitzer high-resolution spectrometer: 
SH 9.6-19.5$\mu$m with slit size 4.7$\arcsec$ $\times$ 11.3$\arcsec$ and LH 19-39$\mu$m 
with slit size 11.1$\arcsec$ $\times$ 22.3 $\arcsec$
allow multi-aperture photometry in the overlapping parts (19.0-19.5$\mu$m).
The ratio of the flux measured in LH to that measured in SH gives the extendedness of the source.
We used this measure of the extendedness of the source to estimate the line emitting regions \citep{t08}.
In Figure 2a we plot the [NeII]12.8$\mu$m line equivalent width as a function of the source extendedness.
We notice that those sources showing a significant mid-IR extendedness 
are mostly type 2 objects or non-Seyfert galaxies and have the highest 
[NeII]12.8$\mu$m line equivalent width, with only two exceptions of very nearby Seyfert 1's. 
An high [NeII]12.8$\mu$m line equivalent width is
a measure of a strong star formation component. This is not the case for the high excitation lines, 
originated from the AGN, such as [NeV], for which no apparent trend appears between 
source extendedness and line EWs , as can be seen in Figure 2b \citep{t08}.


\section{Semianalytical models to estimate AGN and Starburst components }


To estimate the percentage of the contribution of AGN and Starburst in the observed emission 
in a galaxy at 19$\mu$m, we computed an analytical model, 
for each of the observed quantities as a 
function of R, which is defined as the ratio of the Starburst flux at 19$\mu$m to the AGN flux 
at the same wavelength. R  varies from zero - emission totally from the AGN - to infinity - 
emission totally from the Starburst. The chosen observed quantities are: the extendedness of
the source, the equivalent width of PAH at 11.25$\mu$m and [NeII] at 12.81$\mu$m, the line ratio 
[NeV]14.32$\mu$/[NeII]12.81$\mu$m, the spectral index $\alpha$ at (60-25)$\mu$m.

In Fig.3a and 3b the diagrams of [NeV]/[NeII] vs $\alpha$(60-25)$\mu$ m and EW [NeII] vs EW PAH, 
are shown as examples, where the black curves represent the models. 
By inverting the models, i.e. solving the analytical expressions with the true values 
of the observed quantities, we obtained a value of R for each of the five observed quantities 
of every source. We then computed the mean value of R to estimate the relative percentage 
of AGN and Starburst emissions. 


The sample of sources to which we apply this analysis is reduced because the models depend on 
the extendedness and the PAH equivalent width, which are not detected (for the PAH) or measurable 
(for the extendedness) for all the sources(from 86 to 59 objects). 
We found that the model can disentangle the AGN and the Starburst emission for 16 
of 20 Seyfert 1's, 14 of 16 HBLR's, 13 of 14 nonHBLR's, 7 of 9 objects re-classified 
as not Seyferts. The details of this approach will appear in Tommasin et al. (2009).


\begin{figure}[!t]
\plottwo{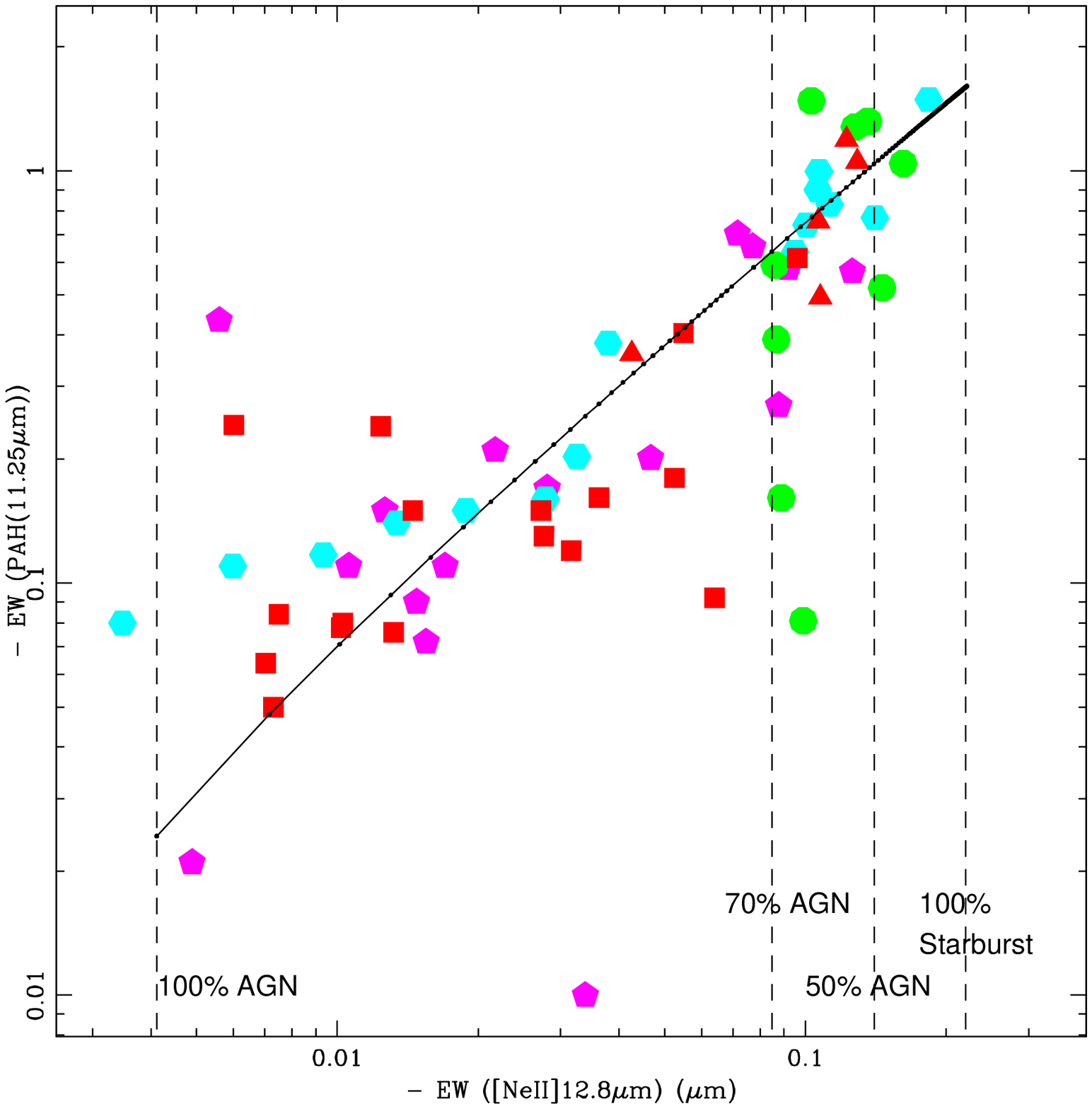}{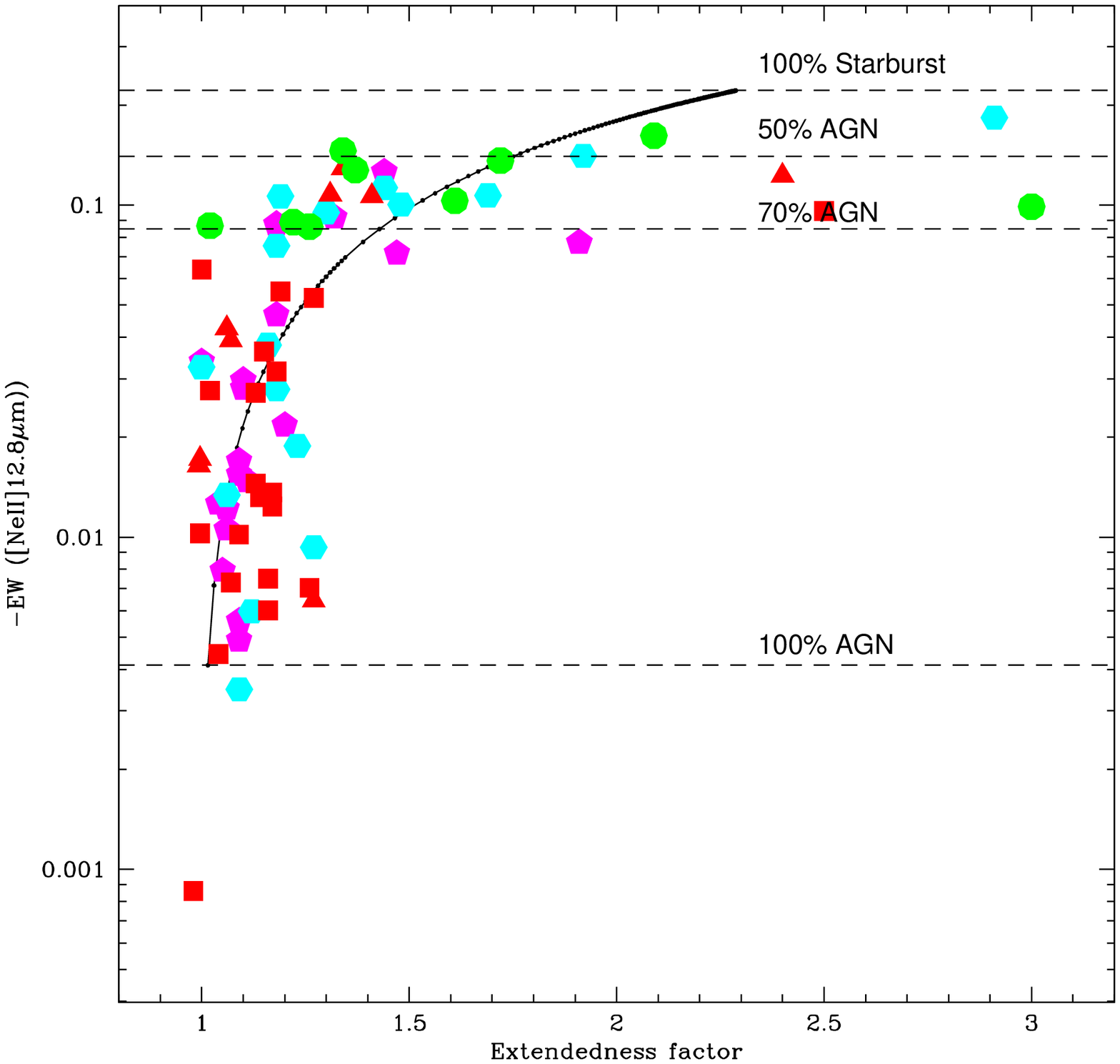}  
    \caption{ {\bf a:(left)} The spectral index $\alpha_{(60\mu m-25\mu m)}$ versus the [NeV]14.3$\mu$m/[NeII]12.8$\mu$m line 
    ratio:  the solid line fitting the data is the model and the dashed lines separate the
diagram in 3 regions with different AGN contribution, (100\% - 70\%),(70\% - 50\%), (50\% - 0\%).
{\bf b:(right)} EW [NeII]12.8$\mu$m vs EW PAH 11.25$\mu$m: same notation as in Fig.3a. }
  \label{fig:kevin}
\end{figure}




\section{Analysis of the 12$\mu$m sample multi-frequency dataset}
\label{sec:ana}

\begin{figure}[!t]
\plottwo{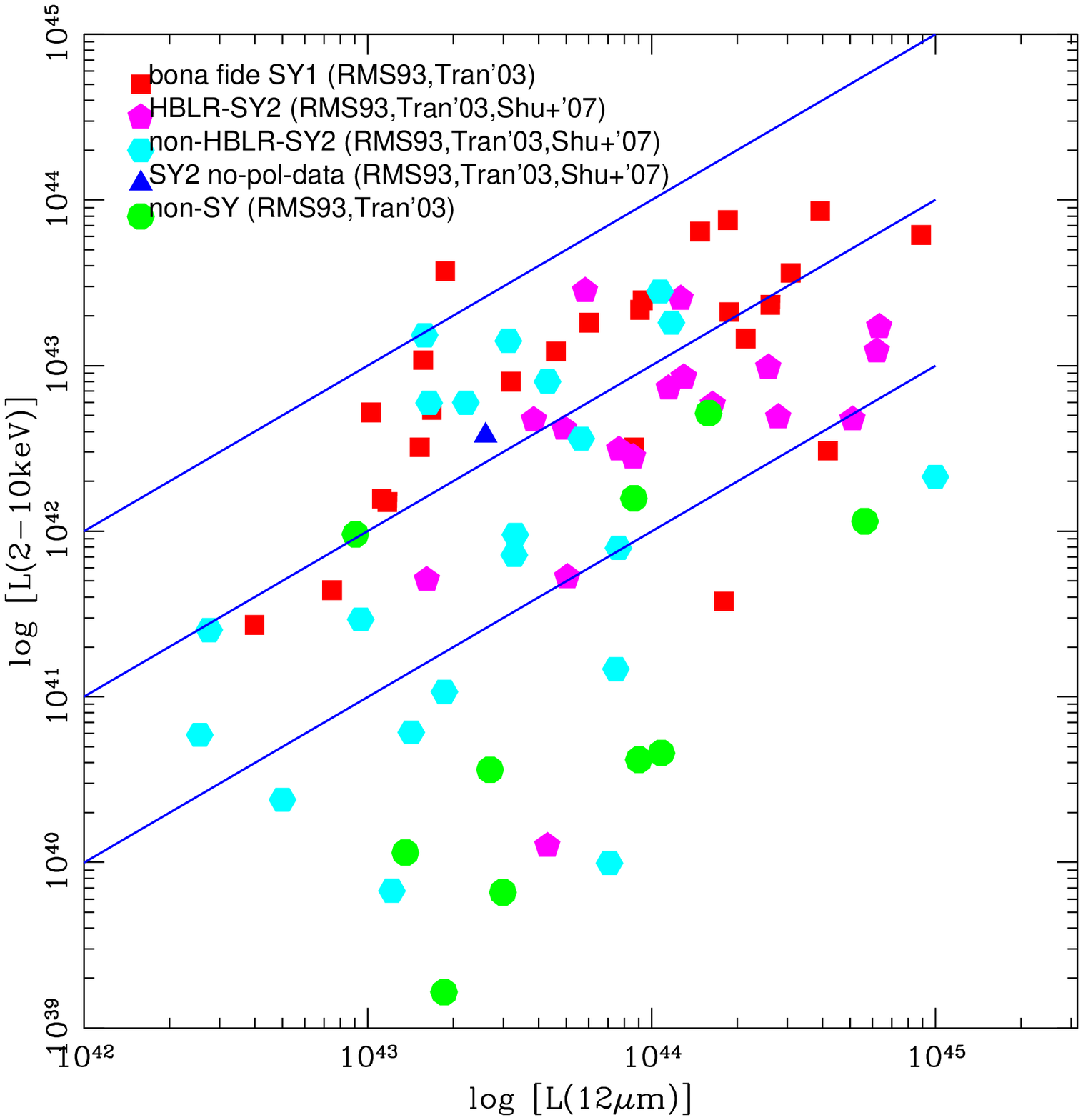}{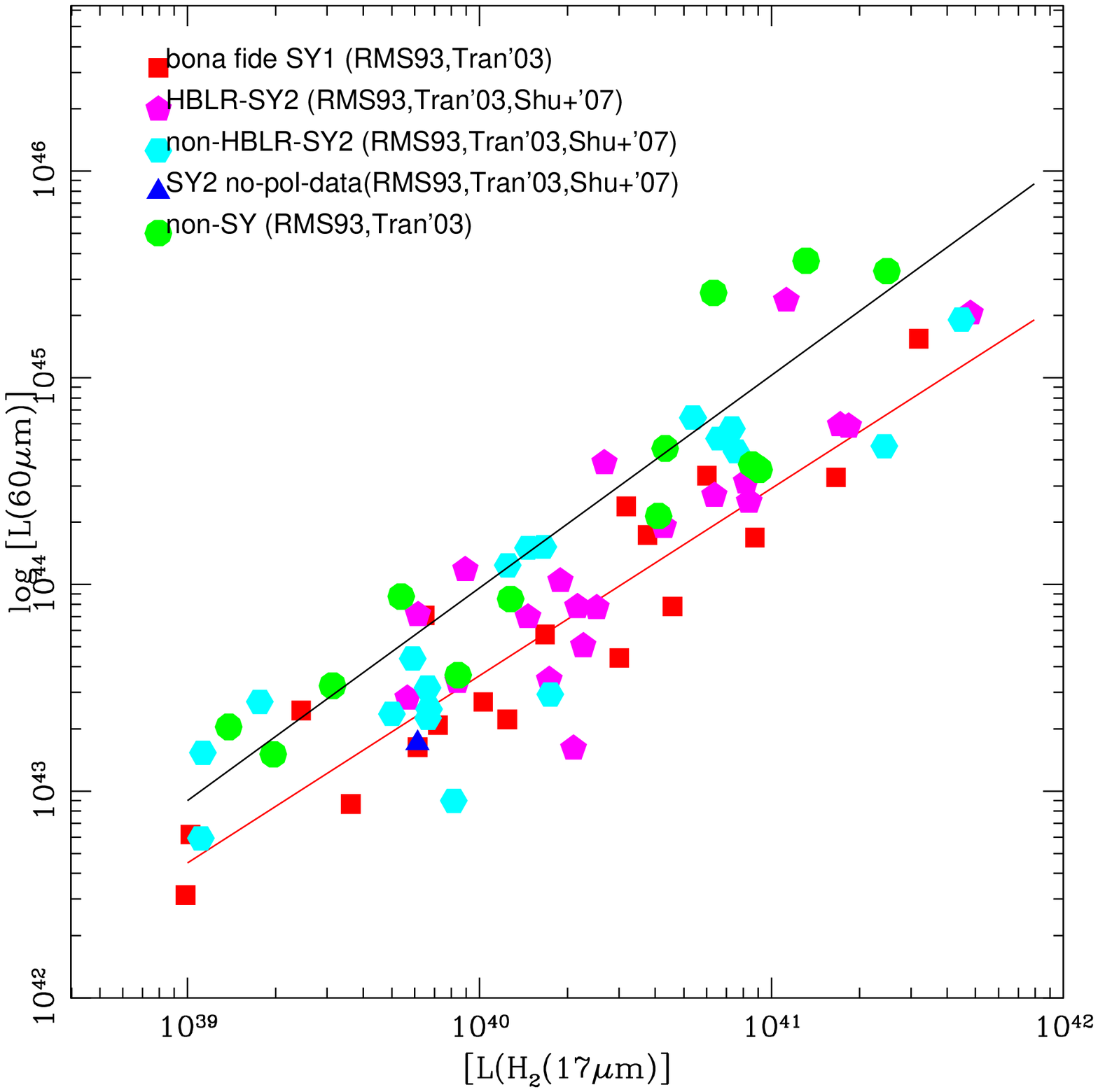}
  \caption{ {\bf a:(left)} Corrected (unabsorbed) X-ray (2-10 keV) luminosity as a function of the 12$\mu$m luminosity. 
  The three lines
  from the top to the bottom indicate the loci of L(2-10keV) = L(12$\mu$m) (\textit{upper});  
  L(2-10keV) = 0.1 $\times$ L(12$\mu$m) (\textit{middle}); and L(2-10keV) = 0.01 $\times$ L(12$\mu$m) 
  (\textit{lower}), which are used in the text to roughly separate the different objects.  {\bf b:(right)} 
  Total 60$\mu$m luminosity as a function of the H$_2$ 17$\mu$m line luminosity.
  The two lines from the top to the bottom are least squares fits of the non-Seyfert galaxies and of the 
  Seyfert 1's, respectively (see the text)}
  \label{fig:lx_l12}
\end{figure}

The 12MGS has been observed extensively from the radio to the X-rays and we can use the large set
of data to search for correlations between different observed quantities. 
To show an example, we want to relate the X-ray luminosity, measuring the accretion, to the bolometric
luminosity, as given by the  12$\mu$m luminosity.
We plot in Figure~\ref{fig:lx_l12} 
the \textit{unabsorbed} 2-10keV luminosity and the 12$\mu$m luminosity.

Following the finding of Spinoglio \& Malkan (1989) and Spinoglio et al (1995) 
that the 12$\mu$m luminosity is
linearly proportional to the \textit{bolometric} luminosity, at a given L$_{bol}$ 
in Figure 4a a sequence can be identified with decreasing accretion luminosity: from 
Seyfert 1's $\rightarrow$ HBLR-Seyfert 2's  $\rightarrow$ \textit{pure} Seyfert 2's.
Although these results are to be considered preliminary, as no statistical method has
yet been applied, 
most Seyfert 1's have: 

$0.1 \times L(12{\mu}m) < L(2-10keV) < L(12{\mu}m)$

\noindent Most HBLR-Seyfert 2's have: 

$0.01 \times L(12{\mu}m) < L(2-10keV) < 0.1 \times L(12{\mu}m)$

\noindent Most pure Seyfert 2's and non-Seyfert's have: 

$L(2-10keV) < 0.01 \times L(12{\mu}m)$

We preminilarily suggest that black hole accretion, as measured by X-rays, is the dominant mechanism
determining the observational nature of a galaxy: when accretion is not an important energy source, we
have galaxies without Seyfert nuclei, dominated by stellar evolution processes (called here non-Seyfert's), 
then when accretion increases we have a sequence from
the \textit{pure} Seyfert 2's, to the HBLR-Seyfert 2's and finally when  accretion dominates the bolometric luminosity,
we have the Seyfert 1's.

In an analogous way, we try to correlate the IRAS 60$\mu$m luminosity (measuring the integrated star formation activity) 
and the H$_2$ S(1) line luminosity (typical star formation indicator) in Figure 4b

\noindent Most Seyfert 1's and HBLR-Seyfert 2's have: 

$L(H_2) \sim 5 \times 10^{-4} \times L(60{\mu}m)$

\noindent Most pure Seyfert 2's and non-Seyfert have: 

$L(H_2) \sim 10^{-4} \times L(60{\mu}m)$


If we make least squares fits to the two extreme populations of Seyfert type 1's and non-Seyfert
galaxies, we obtain a sequence of two almost parallel 
lines of the form $Log(L(60{\mu}m)) = a \times log(L(H_2)) + b $ from the bottom to the top: 
\begin{itemize}
\item[-] for Seyfert 1's: $a$=0.905, $b$=7.325, with a regression coefficient of R=0.928;
\item[-] for non-Seyfert's: $a$=1.030, $b$=2.797, with R=0.925.
\end{itemize}

There are two interpretations of this behavior: either the more active galaxies (type 1's) have enhanced
H$_2$ emission \citep{rig02}, or at a given H$_2$ luminosity, type 2's (and non-Sy) have $L(60{\mu}m)$ 5 
times higher than type 1's, because of an enhanced star formation process.

\end{document}